\documentclass[a4paper]{panl}
\usepackage{cite}
\usepackage{wrapfig}
\usepackage{graphicx}
\usepackage{amssymb}
\usepackage{amsfonts}
\usepackage{amsmath}
\usepackage{longtable}
\usepackage{rotating}
\usepackage{lscape}
\usepackage{epsfig}
\usepackage{xcolor}
\usepackage{graphicx,epstopdf}
\usepackage{tikz}
\usepackage{lineno} 

\originalTeX
\begin{document}
\issuearea{Physics of Elementary Particles and Atomic Nuclei. Theory}

\title{Trigger efficiencies of a proposed beam monitoring detector~(BeBe) for p+p collisions at NICA energies}
\maketitle
\authors{M. A.\,Ayala-Torres$^{a,b,c,}$\footnote{E-mail: ayalatorresm@gmail.com ~~~~~~ \textcolor{blue}{Submitted to PEPAN letters (AYSS2022)}}, L. G.\,Espinoza Beltr\'an$^{d,e}$, L. A.\,Hern\'andez-Cruz$^{d}$,}
\authors{L. M. \,Monta\~no$^{a}$, E.\,Moreno-Barbosa$^{d}$,
and C. H.\,Zepeda Fern\'andez$^{d,f}$}

\from{$^a$\,Departamento de Física, Centro de Investigación y de Estudios Avanzados del Instituto Politécnico Nacional (Cinvestav), 14-740 07000, Mexico City, Mexico}
\from{$^b$\,Millennium Institute for Subatomic physics at high energy frontier-SAPHIR, Fernandez Concha 700, Santiago, Chile} 
\from{$^c$\,Center for Theoretical and Experimental Particle Physics, Facultad de Ciencias Exactas, Universidad Andres Bello, Fernandez Concha 700, Santiago, Chile} 
\from{$^d$\,Facultad de Ciencias Físico Matemáticas, Benemérita Universidad Autónoma de Puebla, Av. San Claudio y 18 Sur, Edif. EMA3-231, Ciudad Universitaria 72570, Puebla, M\'exico}
\from{$^e$\,Facultad de Ciencias F\'isico-Matem\'aticas, Universidad Aut\'onoma de Sinaloa, Avenida de las Am\'ericas y Boulevard C.P. 80000, Culiac\'an, Sinaloa, M\'exico}
\from{$^f$\,C\'atedra CONACyT, 03940, CdMx M\'exico}

\begin{abstract}
The Multipurpose Detector (MPD) consists of a typical array of sub-detectors to study the nuclear matter originating from the collisions of beams provided by the Nuclotron-based Ion Collider fAcility (NICA). A beam monitoring detector~(BeBe) is proposed for stage 2 of MPD to increase the trigger capabilities. BeBe is constituted of two plastic scintillator disks segmented in 80 cells $\pm~2~m$ away from the interaction point of MPD. Laboratory measurements to obtain the energy resolution of an individual BeBe cell prototype are presented. It is shown that an energy resolution of $22\pm6\%$ can be obtained. Based on Monte Carlo simulations, the trigger efficiencies of the BeBe are presented for p+p collisions at 11~GeV considering a threshold in the energy loss of the charged particles reaching the detector.
\end{abstract}
\vspace*{6pt}

\noindent

\section{Introduction}
\begin{figure}[t!]
\begin{center}
\includegraphics[width=.95\textwidth]{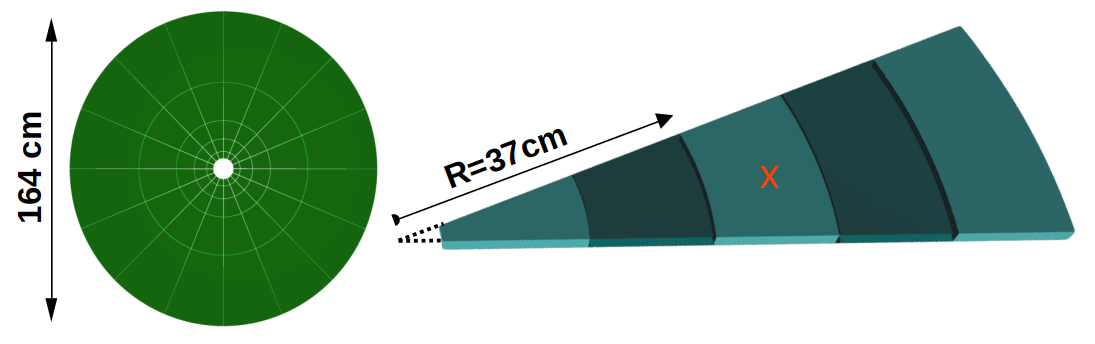}
\vspace{-3mm}
\caption{Left) The geometry of the plastic scintillation disc constituting one of the two BeBe identical hodoscopes. Right) One segment from the disk. The geometry of the prototype used for the laboratory measurements is from the third ring.}
\end{center}
\label{fig:BeBe}
\vspace{-5mm}
\end{figure}

The quantum chromodynamics~(QCD) phase diagram was explored in certain regions of the parameter space by different experiments and a critical endpoint in this diagram is a theory-based prediction. In the Nuclotron-based Ion Collider fAcility (NICA) the Multipurpose Detector (MPD) is currently under construction intending to confirm this prediction~\cite{Golovatyuk:2019rkb,MPD:2022qhn}. The beams provided by NICA in the first stages corresponds to Bi+Bi, Au+Au, and p+p. 
Motivated by the low trigger efficiency in low multiplicity p+p collision events given by the Fast Forward Detector (FFD) of MPD, a complementary detector is proposed (BeBe)~\cite{Torres:2021jgh}. BeBe is constituted of two hodoscopes, each conformed by 80 plastic scintillator cells $\pm~2~m$ away from the interaction point of MPD (see figure~\ref{fig:BeBe}). 
Laboratory measurements to obtain the energy resolution of an individual BeBe cell prototype are presented. Based on Monte Carlo simulations, considering a minimum value for the energy loss, the trigger efficiencies of the BeBe detector for p+p collisions at 11~GeV are presented.

\section{Laboratory measurements}
\label{sec:setup}

\begin{figure}[hbt!]
\begin{center}
\includegraphics[width=.9\textwidth]{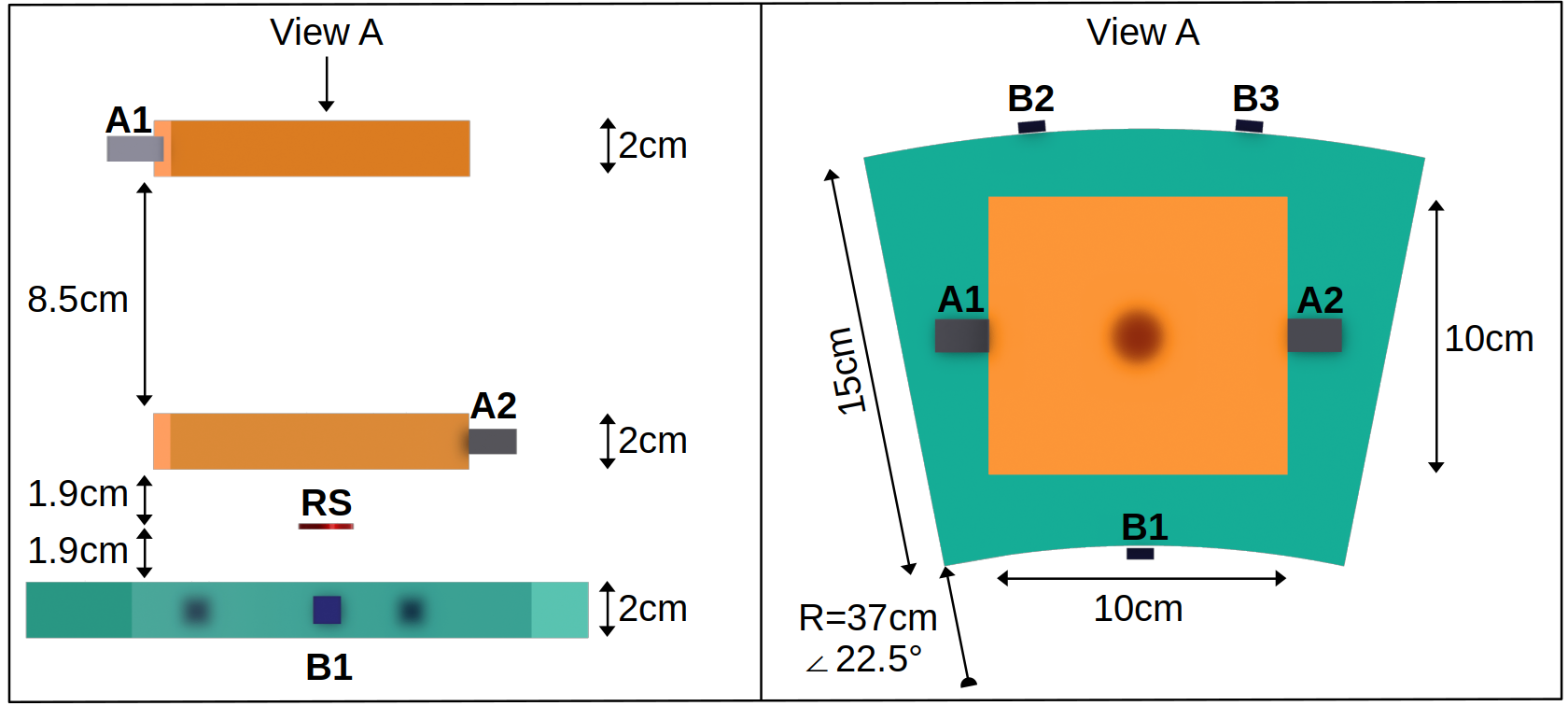}
\vspace{-3mm}
\caption{Diagram of the displacement of the counters~(plastic scintillator in yellow and the PMT in gray), the prototype~(plastic scintillator in green and the 3 SiPMs in blue), and the radioactive source~(in red).}
\end{center}
\label{fig:setup}
\vspace{-5mm}
\end{figure}

The counters used in the experimental setup consist of an array of photodetectors coupled to ultra-fast plastic scintillator BC-404 from Saint-Gobain Crystals wrapped with one layer of Tyvek and two layers of Mylar. 
The experimental setup is shown in figure~\ref{fig:setup}. An array of counters coupled to a photomultiplier tube-H5783~(PMT, A1 and A2) was used to implement the logic used to digitize the pulse obtained from the 3 silicon photomultipliers-S13360-3050CS~(SiPMs, B1-B3)
attached to the prototype (in green). A radioactive source~(RS, in red) made of $^{22}Na$ in the form of a thin disk was placed between A2 and the prototype. This RS decay via $\beta^+$ and annihilation radiation ($\gamma$ with energy of 0.511~MeV) via positron capture can be measured from a coincidence AND between A2 and the prototype. The main background of our measurement comes from secondary particles of cosmic rays. To be more selective, therefore, we choose the events from a coincidence from A2 AND the prototype but not from A1~(see figure 3). Under this implemented logic, we performed measurements with and without the radioactive source. 

The threshold values for the selection criteria correspond to 10, 10, and 25~ADC units for the signals from A1, A2, and the prototype, respectively. The resulting amplitude distributions from the prototype are shown in figure~4. The signal from the RS corresponds to the peak centered at 25~mV. The RS signal distribution was obtained considering the bin-by-bin difference between the data with RS minus the data without RS in the signal region (15-35~mV). The RS signal distribution has an FWHM=$5.3\pm1.4$~mV and mean=$25.1\pm1.5$~mV. Then, the energy resolution of the prototype is $\sigma=\frac{FWHM}{mean}=22\pm6\%$.  

\begin{figure}[hbt!]
\begin{center}
\includegraphics[width=.55\textwidth]{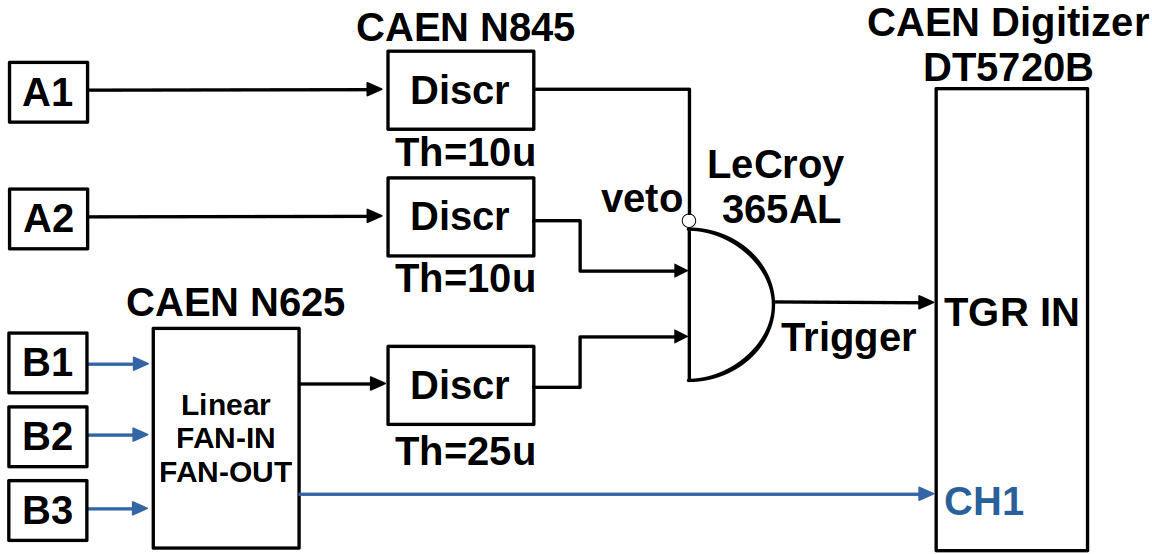}
\vspace{-3mm}
\caption{Electronic logic used to digitize the pulse from the prototype. A coincidence signal from A2 and the prototype with a veto from A1 provide the start signal for the data readout.}
\end{center}
\label{fig:DAQ}
\vspace{-5mm}
\end{figure}

\begin{figure}[hbt!]
\begin{center}
\includegraphics[width=.95\textwidth]{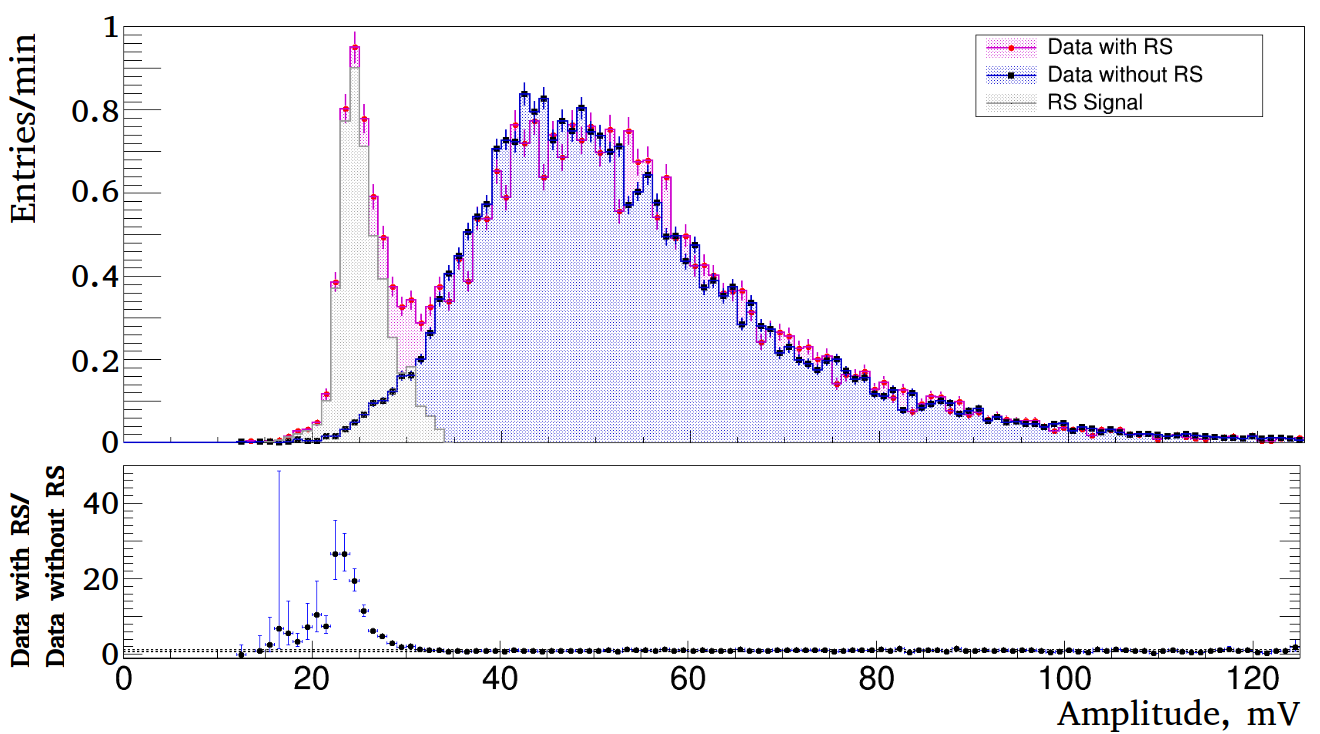}
\vspace{-3mm}
\caption{Amplitude distributions from the prototype with~(magenta) and without~(blue) the radioactive source~(RS) and the difference (with - without) between these  distributions is shown in gray. The ratio plot with/without RS is shown at the bottom panel in which the distinction between background and signal is clear.}
\end{center}
\label{fig:Amplitude}
\vspace{-5mm}
\end{figure}

\newpage
\section{BeBe trigger efficiencies}
The main purpose of the BeBe detector is to be a luminometer in the MPD-NICA experiment. The trigger signal of the BeBe generated by either one hit in any of the two BeBe arrays located at opposite sides of the MPD-NICA experiment or two hits in coincidence at a certain time window in both of them can be used for the luminosity determination of the NICA beam, like VZERO-ALICE~\cite{Aamodt:2008zz} and LUCID-ATLAS~\cite{Avoni:2018iuv} detectors at LHC.

We considered one reduced geometry with the closest proposed detectors to the MPD interaction point: Mbb, FFD, and BeBe detectors. Also, we do not consider smearing in the beam. We use a sample of p+p collisions at 11~GeV with 1~million events generated using UrQMD Monte Carlo model ~\cite{Bass:1998ca,Bleicher:1999xi} within MPDRoot framework~\cite{MPDROOT}.

We defined a time window of 4 to 7 ns to the time of flight distribution for each matrix to simulate the following BeBe trigger flags: \\
$\bullet$ BBR/BBL: if the Z coordinate of the BeBe hit is positive/negative. \\
$\bullet$ BBR AND BBL: logical AND of the coincidence of BBR and BBL. \\
$\bullet$ BBR OR BBL: logical OR of BBR and BBL. \\
Figure~5 shows the energy loss and the number of particles reaching the BeBe rings event-by-event before and after the cut in the energy deposition of 4~MeV was applied. After this cut, the number of particles reaching BeBe is reduced by 24\%. 
Table~\ref{tab:TriggerEff} summarises the results obtained for each trigger flag.  The flag BBR OR BBL shows a reduction of 4\% with respect to the trigger efficiency that can be obtained at the level of particle hit.
\begin{figure}[t!]
\begin{center}
\includegraphics[width=.49\textwidth]{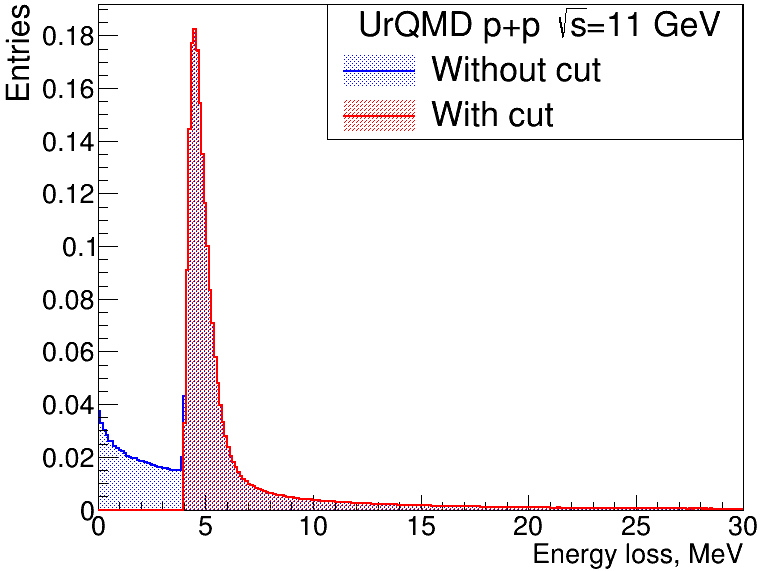}
\includegraphics[width=.49\textwidth]{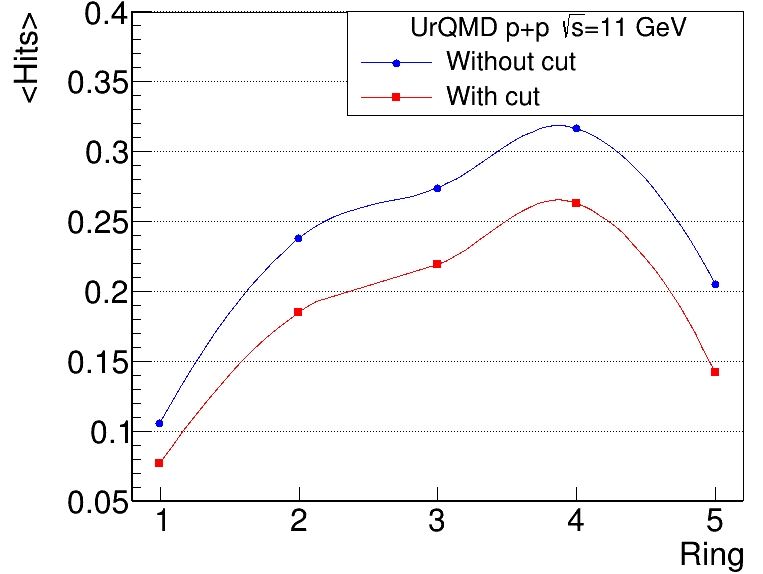}
\vspace{-3mm}
\caption{
Left) The ionizing energy loss in the plastic cells of BeBe and 
Right) The average charged particles per ring with and without a cut in the energy deposited per particle in the BeBe cells. All the distributions are normalized to the number of simulated events. 
}
\end{center}
\label{fig:PartRing}
\vspace{-10mm}
\end{figure}

\begin{table}[t!]
\centering
\begin{tabular}{c c c c c}
\hline \hline
BBR & BBL & BBR and BBL & BBR or BBL & Energy deposited cut \\\hline
74.31\% & 74.42\% & 52.7\% & 96.03\% & No \\ \hline
69.39\% & 69.39\% & 46.67\% & 92.13\% & Yes \\
\hline\hline
\end{tabular} 
\caption{Trigger efficiencies of BeBe for the process p+p at 11~GeV without smearing in the vertex. Simulated detectors Mbb, FFD, and BeBe. The first row of the table was taken from~\cite{Torres:2021jgh}.
}
\label{tab:TriggerEff}
\end{table}

\newpage
\section{Conclusions}
We measured the energy resolution of an individual BeBe cell equal to $22\pm6\%$ for $\gamma$ rays with an incident energy of 0.511~MeV. 
Considering the trigger flag OR between the two hodocopes of the simulated BeBe detector, a trigger efficiency for p+p collisions at 11~GeV of 92\% is reached. 

\section*{Acknowledgments}

The authors thank Eng. Marcos Fontaine for helping with the SiPM electronics.
This work was supported by CONACyT research grants A1-S-23238 and A1-S-13525, and ANID Millennium Program ICN2019 044.
L.G.E.B. acknowledges support from postdoctoral fellowships granted by Consejo Nacional de Ciencia y Tecnolog\'ia. M.A.A.T. acknowledges support from ANID postdoctoral grant 3230806.



\end{document}